\documentclass[twocolumn,amsfonts,amssymb,amsmath,superscriptaddress]{revtex4-1}

\usepackage{graphicx} %
\usepackage{soul} %
\graphicspath{{figs/}} %
\usepackage[colorlinks = true, %
            linkcolor = blue,
            urlcolor  = blue,
            citecolor = blue,
            anchorcolor = blue]{hyperref}
\usepackage[dvipsnames]{xcolor} %
\usepackage[T1]{fontenc}
\usepackage[utf8]{inputenc}
\usepackage[normalem]{ulem}

\begin{document}
\title{Phonon collapse and anharmonic melting of the 3D charge-density wave in kagome metals}

\author{Martin Gutierrez-Amigo}
\thanks{Corresponding author - Martin Gutierrez-Amigo \\
amigo.martin@ehu.eus
}

\affiliation{Donostia International Physics Center (DIPC), 20018 Donostia/San Sebasti\'an, Spain}
\affiliation{Department of Physics, University of the Basque Country (UPV/EHU), 48080 Bilbao, Spain}
\affiliation{Centro de Física de Materiales (CSIC-UPV/EHU), 20018 Donostia/San Sebasti\'an, Spain}

\author{{\DJ}or{\dj}e Dangi{\'c}}
\affiliation{Centro de Física de Materiales (CSIC-UPV/EHU), 20018 Donostia/San Sebasti\'an, Spain}
\affiliation{Fisika Aplikatua Saila, Gipuzkoako Ingeniaritza Eskola, University of the Basque Country (UPV/EHU), 20018 Donostia/San Sebasti\'an, Spain}

\author{Chunyu Guo}
\affiliation{Max Planck Institute for the Structure and Dynamics of Matter, 22761 Hamburg, Germany}

\author{Claudia Felser}
\affiliation{Max Planck Institute for Chemical Physics of Solids, 01187 Dresden, Germany}

\author{Philip J. W. Moll}
\affiliation{Max Planck Institute for the Structure and Dynamics of Matter, 22761 Hamburg, Germany}

\author{Maia G. Vergniory}
\affiliation{Max Planck Institute for Chemical Physics of Solids, 01187 Dresden, Germany}
\affiliation{Donostia International Physics Center (DIPC), 20018 Donostia/San Sebasti\'an, Spain}

\author{Ion Errea}
\affiliation{Centro de Física de Materiales (CSIC-UPV/EHU), 20018 Donostia/San Sebasti\'an, Spain}
\affiliation{Fisika Aplikatua Saila, Gipuzkoako Ingeniaritza Eskola, University of the Basque Country (UPV/EHU), 20018 Donostia/San Sebasti\'an, Spain}
\affiliation{Donostia International Physics Center (DIPC), 20018 Donostia/San Sebasti\'an, Spain}

\date{November 22, 2023}
\begin{abstract}
The charge-density wave (CDW) mechanism and resulting structure of the $\mathrm{AV_{3}Sb_{5}}$ family of kagome metals has posed a puzzling challenge since their discovery four years ago.
In fact, the lack of consensus on the origin and structure of the CDW hinders the understanding of the emerging phenomena.
Here, by employing a non-perturbative treatment of anharmonicity from first-principles calculations, we reveal that the charge-density transition in $\mathrm{CsV_{3}Sb_{5}}$ is driven by the large electron-phonon coupling of the material and that the  melting of the CDW state is attributed to ionic entropy and lattice anharmonicity.
The calculated transition temperature is in very good agreement with experiments, implying that soft mode physics are at the core of the charge-density wave transition.
Contrary to the standard assumption associated with a pure kagome lattice, the CDW is essentially three-dimensional as it is triggered by an unstable phonon at the $L$ point. 
The absence of involvement of phonons at the $M$ point enables us to constrain the resulting symmetries to six possible space groups.
The unusually large electron-phonon linewidth of the soft mode explains why inelastic scattering experiments did not observe any softened phonon.
We foresee that large anharmonic effects are ubiquitous and could be fundamental to understand the observed phenomena also in other kagome families.

\end{abstract}
\maketitle
\newpage

\begin{figure*}[]
		\centering
		\includegraphics[width=\textwidth]{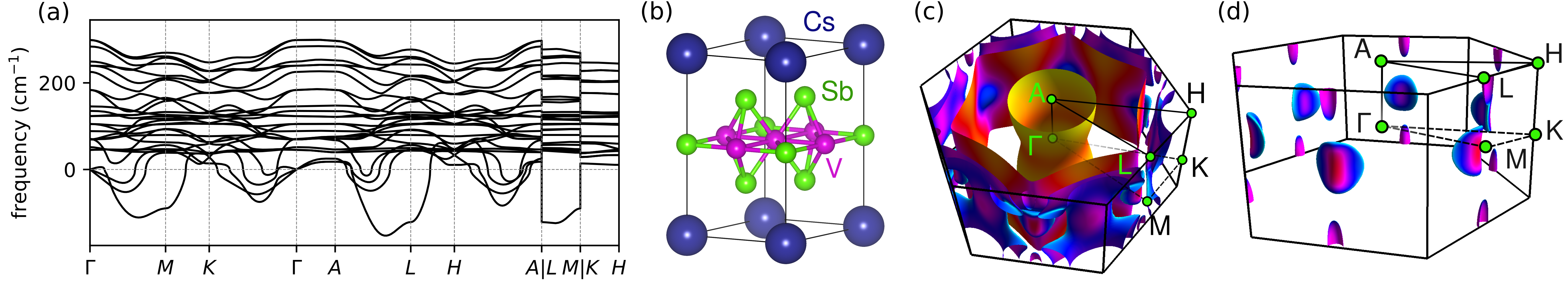}
		\caption{
		\textbf{Harmonic phonons and Fermi suface  in the high-symmetry phase of $\mathbf{CsV_3Sb_5}$.} 
		\textbf{a.} Calculated harmonic phonon dispersion of $\mathrm{CsV_3Sb_5}$ in the $P6/mmm$ phase. The harmonic spectrum exhibits two main instabilities at the $M$ and $L$ high-symmetry points. 
		\textbf{b.} Unit cell for $\mathrm{CsV_3Sb_5}$. The vanadium atoms (pink) form a perfect kagome lattice.
		\textbf{c.} Fermi surface within the first Brillouin zone with labeled high-symmetry points.
		\textbf{d.} While most of the Fermi surface exhibits low dispersion in the $z$ direction, there are closed Fermi surface pockets containing Van Hove singularities situated near the $M$ and $L$ points, indicating the three-dimensional nature of $\mathrm{CsV_3Sb_5}$.
		}
		\label{fig:1}
\end{figure*}

\section{Introduction}
The kagome lattice, composed of three triangular lattices rotated 120 degrees with respect to each other, stands as one of the most thoroughly investigated hexagonal lattices, offering a wealth of intriguing electronic properties linked to its distinct geometry \cite{syozi_1951}.
Its multilattice nature allows for the emergence of flat bands \cite{calugaru_2022,kang_2020,kang_2020a,bilitewski_2018}, which, in turn, lead to high electronic correlation effects provided they fall into the vicinity of the Fermi energy \cite{ghimire_2020,mazin_2014,zhao_2021,guo_2023},
while its triangular arrangement provides a natural platform for magnetically frustrated systems \cite{sachdev_1992,kang_2020}.
The presence of Dirac cones in the band structure leads also to non-trivial topological effects \cite{yin_2022,ghimire_2020}.
Consequently, the recently discovered family of kagome materials, $\mathrm{AV_{3}Sb_{5}}$ with $\mathrm{A={Cs,K,Rb}}$ \cite{ortiz_2019}, has garnered significant attention, as it provides a platform for exploring the interplay between electronic correlations, frustrated geometry, charge-density waves (CDWs), topology, and even superconductivity.

All compounds within the $\mathrm{AV_{3}Sb_{5}}$ family exhibit a CDW at a temperature of approximately $90\ \mathrm{K}$ \cite{ortiz_2019}. The
CDW holds a crucial significance because  below $T_{CDW}$ a plethora of novel and intriguing physical phenomena have been reported. These include switchable chiral transport \cite{guo_2022,guo_2023a}, specular optical rotation \cite{farhang_2023}, or the presence of a chiral flux phase \cite{yu_2021,feng_2021,mielke_2022} accompanied by loop currents.
This unconventional nature is also manifested in the superconducting order observed below 2.5 K for $\mathrm{CsV_{3}Sb_{5}}$ \cite{ortiz_2020}, with reports, for instance, of unconventional superconductivity \cite{tan_2021,chen_2021b,zhao_2021a,wu_2021}, multi-charge flux quantization \cite{ge_2022}, and chiral superconductivity \cite{le_2023}.

The true nature of the CDW and the resulting atomic structure remain open questions.
Moreover, it is not clear what the intricate relation between the CDW order and all observed unconventional phenomena is. 
In fact, the absence of consensus regarding the CDW structure hinders the understanding of the emerging properties, mainly because they might be constrained by symmetry.
Just below $T_{CDW}\sim 94\ \mathrm{K}$, the prevailing experimental evidence supports a three-dimensional $2\times 2\times 2$ structure for the CDW \cite{ratcliff_2021,li_2021,liu_2022a,subires_2023}, but without a consensus on whether the CDW breaks six-fold symmetry \cite{zhao_2021,ratcliff_2021,xu_2022,nie_2022,kang_2023}.
Some works report a second CDW around $T_{CDW_2}\sim 60\ \mathrm{K}$ \cite{zhao_2021, ratcliff_2021, yu_2021, li_2021, nie_2022, he_2023, stahl_2022}, in which a $2\times 2\times 4$ CDW \cite{ortiz_2021, broyles_2022}, a mixture of $2\times 2\times 2$ with $2\times 2\times 4$ orders \cite{xiao_2023}, or a transition between both types of ordering \cite{stahl_2022} have been reported.
The emergence of the second CDW seems to coincide with the onset of unconventional phenomena, including the chiral flux phase \cite{yu_2021}, activated chiral transport \cite{guo_2022}, and the disruption of $C_6$ symmetry \cite{zhao_2021,nie_2022}.
It is important to note that there are also reports which do not observe this second CDW \cite{liu_2022a}. In fact, it is not clear whether this exotic phenomena is intrinsic to the material or whether it is imposed by external perturbations \cite{guo_2023}.
Doubts also persist on the conservation of time-reversal symmetry, with contradictory results from muon spin spectroscopy \cite{yu_2021,khasanov_2022,shan_2022} and magneto-optical Kerr effect \cite{saykin_2023,xu_2022,farhang_2023}.

The origin and character of the CDW also remains a subject of debate.
At first glance, the nesting mechanism \cite{li_2021,christensen_2021,tan_2021,jiang_2021,denner_2021,lin_2021,jin_2022,deng_2023} appears natural, aligning perfectly with the fermiology of the pure kagome lattice and resulting in the widely experimentally confirmed $2\times2$ modulation within the plane \cite{zhao_2021,ratcliff_2021,yu_2021,li_2021,nie_2022,xie_2022,liu_2022a,subires_2023}.
This nesting paradigm has also been employed to account for the reported unconventional character of the CDW \cite{wang_2021a,jiang_2021,denner_2021,neupert_2022}.
On the other hand, there are also multiple reports suggesting an electron-phonon driven mechanism \cite{xie_2022,liu_2022a,wang_2022}. With regard to the CDW character, reports concur on the absence of observed softening in the phonon spectrum in Raman \cite{ratcliff_2021,liu_2022a} and inelastic x-ray scattering \cite{subires_2023} experiments, suggesting a first-order transition to the CDW.
Even though a discontinuity observed in the lattice parameters at $T_{CDW}$ supports this picture \cite{frachet_2023}, the discontinuity is so minute that a soft phonon mode driven CDW should not be excluded, which would be consistent with the general mechanism in other CDW materials like transition-metal dichalcogenides \cite{weber_2011,diego_2021,bianco_2020}.
Theoretical calculations performed thus far do not clarify these issues as they are mostly limited to phenomenological models \cite{christensen_2021,denner_2021,lin_2021,jin_2022,deng_2023,ferrari_2022} or to \emph{ab initio} calculations within the standard harmonic approximation for the phonons \cite{tan_2021,ratcliff_2021,subedi_2022}, which is known to break down in CDW systems \cite{diego_2021,bianco_2020,bianco_2019,gutierrez-amigo_2024}.
A recent investigation into the temperature dependence of the phonon spectra does suggest the presence of soft mode physics \cite{ptok_2022}, but lacks the resolution of the specific soft mode triggering the CDW or an in-depth spectral analysis.

In this work, making use of first-principles density-functional theory (DFT) calculations including a non-perturbative treatment for lattice anharmonicity, we show that the CDW instability in $\mathrm{CsV_{3}Sb_{5}}$ is triggered by the softening with decreasing temperature of a phonon mode at the $L$ point. The softening is a consequence of its extraordinary large electron-phonon coupling, discarding pure electronic nesting as the main destabilizing force. Our calculated $T_{CDW}$ is in very good agreement with experiments, demonstrating that the CDW melts due to lattice entropy and that soft mode physics plays a dominant role in the CDW transition. This is consistent with a second-order phase transition as well as with a weak  first-order character. We explain that the phonon softening is not observed experimentally \cite{subires_2023} due to its huge broadening, a consequence of the large electron-phonon and anharmonic interactions of the soft mode.

\section{Results and discussion}
\subsection{The melting of the CDW phase}
The DFT harmonic spectrum of the $\mathrm{CsV_{3}Sb_{5}}$ high-symmetry phase exhibits a significant number of lattice instabilities (Fig. \ref{fig:1}(a)), in agreement with previous works \cite{tan_2021,subedi_2022}.
Two primary instabilities can be observed near the $M$ and $L$ points, coinciding with the nesting vectors of the Van Hove Fermi pockets (Fig. \ref{fig:1}(d)).
The most prominent instability occurs at a specific point along the $AL$ line, which we will refer to as the $AL$ mode.
However, the instabilities in the phonon spectra of $\mathrm{CsV_{3}Sb_{5}}$ are highly sensitive to the electronic temperature used in the DFT calculations to perform integrals over the Brillouin zone \cite{christensen_2021}.
Considering that when increasing the electronic temperature in the calculations the $L$ mode stabilizes after the $AL$ one (see supplementary material), the main instabilities of the system seem to be the soft modes transforming under $L_{2}^{- }$ and $M_{1}^{+ }$ irreducible representations in line with previous findings \cite{tan_2021,subedi_2022,wang_2022}.
Considering that each of these points contributes with three equivalent vectors within the star, we expect the CDW to be described by a six-dimensional order parameter $\mathbf{Q}=\left( M_1,M_2,M_3,L_1,L_2,L_3 \right) $.

\begin{figure}[]
		\centering
		\includegraphics[width=\linewidth]{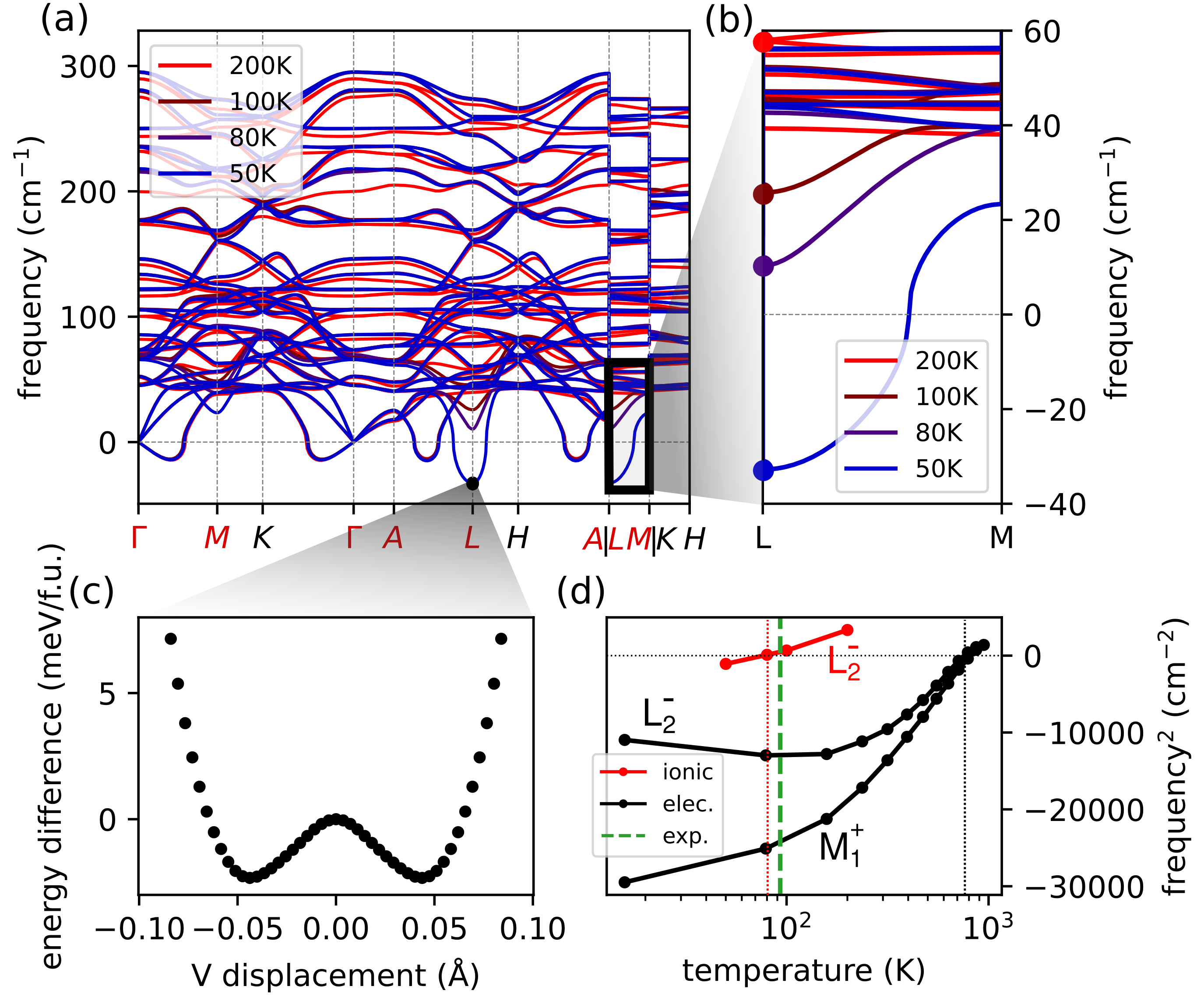}
		\caption{
		\textbf{Anharmonic phonon spectrum of $\mathbf{CsV_3Sb_5}$.}
		\textbf{a.} Calculated anharmonic phonon spectrum for the $P6/mmm$ phase at different temperatures above and below $T_{CDW}\sim 94\ \mathrm{K}$.
		The high-symmetry points labeled in red are explicitly computed, while the rest of the spectrum is interpolated.
		At $80\ \mathrm{K}$ the system is entirely stabilized by anharmonic effects, considering that the remaining imaginary frequencies are artifacts of the Fourier interpolation.
		 \textbf{b.} Zoomed view of the $LM$ path, showcasing the softening of the $L$ mode.
		 Meanwhile, the $M$ phonon remains fully stabilized across the entire temperature range due to anharmonic effects.
		\textbf{c.} Computed Born-Oppenheimer energies as the structure distorts according to the $L_2^{-}$ phonon responsible for the CDW.
		\textbf{d.}
		Effects of electronic and ionic entropy on the stability of the $\mathrm{CsV_3Sb_5}$ high-symmetry phase.
		The black line shows the squared frequency of the $L_2^{-}$ and $M_1^{+}$ modes with respect to electronic temperature, suggesting that the high-symmetry structure is expected to stabilize at around $\sim 815\ \mathrm{K}$.
		Conversely, the red line considers only the ionic entropy for the $L_{2}^{-}$ mode.
		Our calculations predict that ionic entropy stabilizes the system at $\sim 77\ \mathrm{K}$, in agreement with the $T_{CDW}\sim 94\ \mathrm{K}$ experimental value.
		}
		\label{fig:2}
\end{figure}

The high sensitivity of both the $L_{2}^{- }$ and $M_1^{+ }$ modes to the electronic temperature has been used to argue in support of an electron-driven CDW via a nesting mechanism \cite{christensen_2021}.
This concept hinges on the idea that the Van Hove singularities  at   $M$ and $L$  (Fig. \ref{fig:1}(d)) are coupled by the previously described six-dimensional order parameter $\mathbf{Q}$.
As temperature increases, it leads to a reduction in the occupied states associated with these Van Hove singularities, subsequently diminishing the instability.
To test the hypothesis of electronic entropy as a stabilizing factor for the high-symmetry phase, we compute in the harmonic approximation the frequency of the $L_2^{- }$ and $M_1^{+ }$ modes as a function of the real electronic temperature described by Fermi-Dirac statistics.
As shown in Fig. \ref{fig:2}(d), while the modes do eventually stabilize, the predicted transition temperature of approximately $760\ \mathrm{K}$ for the CDW is far from the experimental observations.
This, in conjunction with the highly anharmonic Born-Oppenheimer energy landscape illustrated in Fig. \ref{fig:2}(c), suggests that it is not the electronic entropy, but the ionic entropy, what is responsible for the melting of the CDW as it is the case in transition-metal dichalcogenides \cite{bianco_2020}.

To explore this idea, we compute the static phonon spectra as a function of temperature in the high-symmetry phase of $\mathrm{CsV_{3}Sb_{5}}$, taking into account anharmonic effects within the stochastic self-consistent harmonic approximation (SSCHA) \cite{errea_2014,bianco_2017,monacelli_2018,monacelli_2021}.
As seen in Fig. \ref{fig:2}(a), the anharmonic corrections strongly renormalize the phonon spectrum to the point where it is fully stabilized above $T_{CDW}\sim\ 80\ \mathrm{K}$, which is in very good agreement with the experimental value of 94 K. This good agreement demonstrates that it is ionic entropy, which is largely affected by anharmonicity, what melts the CDW and that electronic entropy does not play any role. 
Interestingly, the $M_1^{+ }$ soft mode is no longer almost degenerate with the $L_2^{- }$ phonon, as it remains stable even at $50\ \mathrm{K}$. This is in agreement with thermal diffuse scattering experiments that do not see any signal at $M$, only at $L$ \cite{subires_2023}.
$\mathrm{CsV_{3}Sb_{5}}$ deviates thus from the ideal two-dimensional kagome instability picture and indicates that the coupling between kagome layers is strong enough to break the degeneracy between the phonons at $M$ and $L$.
This result is in line with quantum oscillation and magnetotransport experiments underscoring the significance of interlayer coupling \cite{huang_2022}. Consequently, a three-dimensional CDW with modulation along the $c$-axis emerges.
This greatly simplifies the analysis of the CDW from a six-dimensional to a three-dimensional order parameter that now is solely related to the $L_{2}^{- }$ instabilities. The clear softening observed at the $L$ point in this static calculation shows that soft phonon physics are triggering the CDW transition and that it has to be of second-order or weak first-order character. 

\begin{figure}[]
		\centering
		\includegraphics[width=\linewidth]{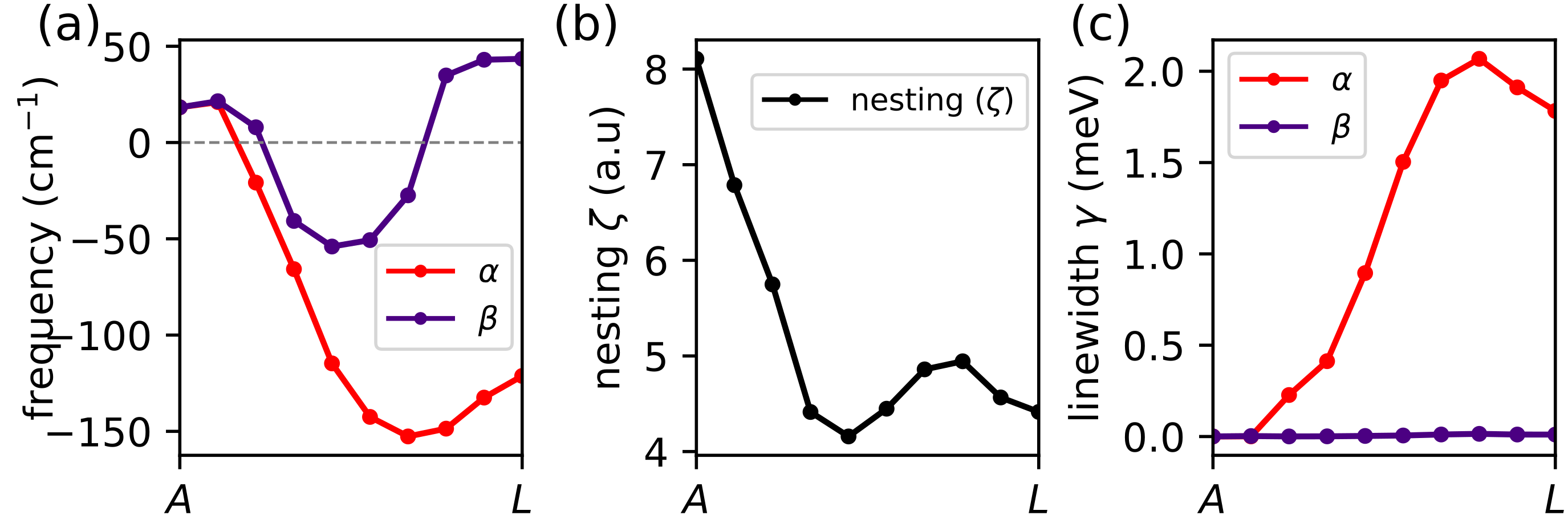}
		\caption{
		\textbf{Nesting and electron-phonon linewidth calculations.}
		All dotted points in the three plots correspond to explicit calculations.
		\textbf{a.} Harmonic phonon frequencies along the $AL$ high-symmetry line for both the $\alpha$ phonon branch (which contains the $L_2^{-}$ instability) and the subsequent most unstable branch, $\beta$.
		\textbf{b.} The nesting function along the $AL$ high-symmetry line exhibits a peak at $A$ rather than at $L$, which contradicts expectations for a nesting-driven mechanism.
		This emphasis on the $A$ point aligns with the highly two-dimensional Fermi surface (Fig. \ref{fig:1}(c)).
		\textbf{c.} The electron-phonon linewidth of the $\alpha$ and $\beta$ branches along the $AL$ line shows a large peak at the $L$ point for the $\alpha$ mode.
		}
		\label{fig:3}
\end{figure}

\begin{figure*}[]
		\centering
		\includegraphics[width=\linewidth]{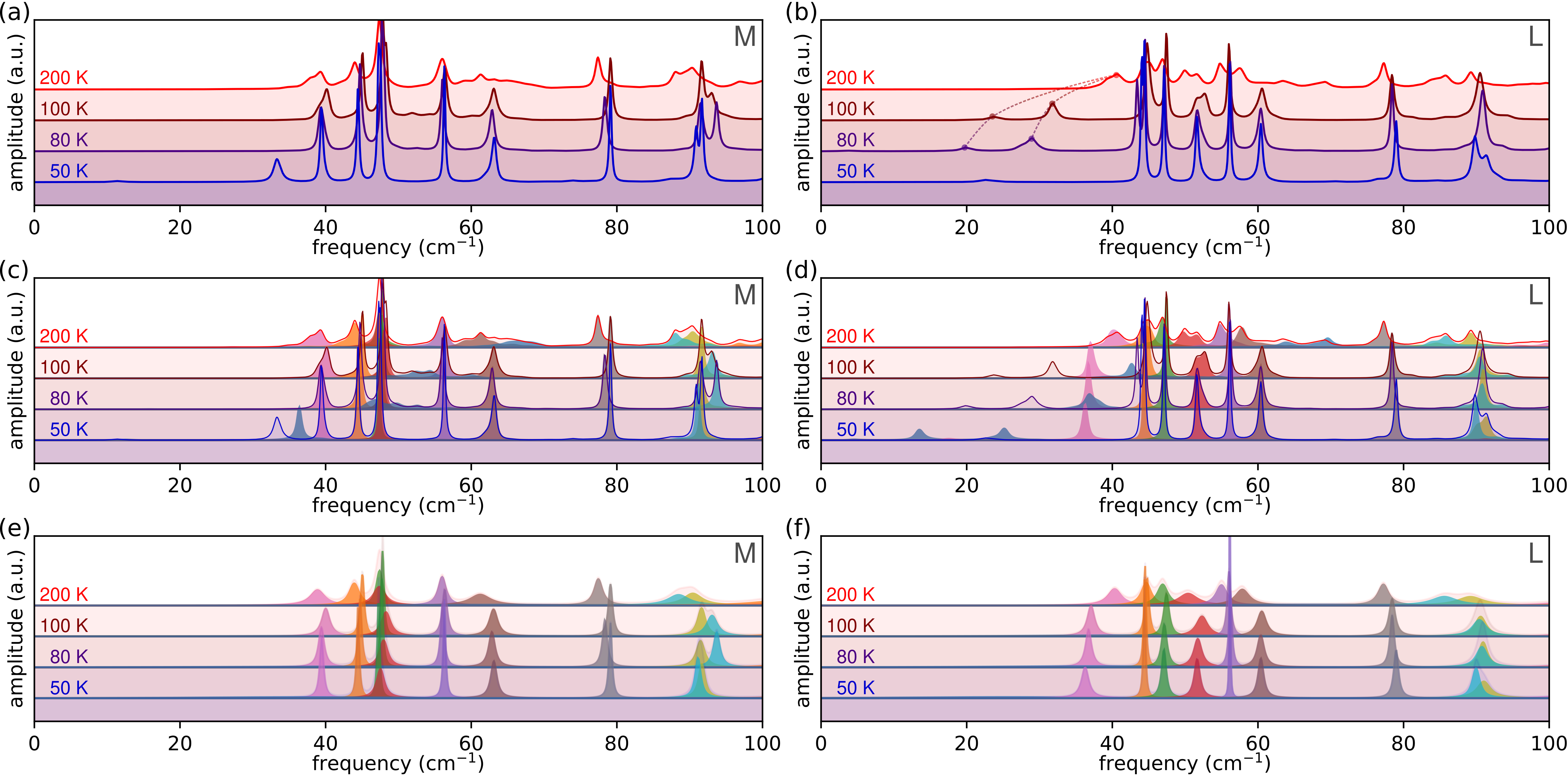}
		\caption{
		\textbf{Calculations of the spectral function at the $M$ and  $L$ points.}
		\textbf{a, b.} Fully anharmonic spectral function for the $M$ and $L$ points at various temperatures.
		While no evident softening is observed at the $M$ point, the $L$ point exhibits a noticeable softening of a peak at approximately 40 $\mathrm{cm}^{-1}$.
		Due to the large anharmonic effects, this peak splits into a double-peak as the temperature lowers.
		\textbf{c, d.} Employing a no-mixing approximation by discarding the off-diagonal elements of the self-energy allows for the independent tracking of different modes, as indicated by different colors.
		The complete anharmonic spectral functions from panels \textbf{a} and \textbf{b} are also overlaid.
		\textbf{e, f.}Here, we fit a Lorentzian function to each of the individual modes in \textbf{c,d} and add the corresponding electron-phonon linewidth to each of the modes.
		}
		\label{fig:4}
\end{figure*}

One important conclusion from our results is that the CDW mechanism is independent of the novel physics that emerge below $T_{CDW}$. In other words, the exotic physics observed are not necessary to explain the CDW.
The study of the low-temperature physics of $\mathrm{CsV_{3}Sb_{5}}$ becomes, then, more feasible.
Instead of dealing with the CDW and the new phenomena in one single problem, one may first solve the CDW structure and then take advantage of the correct symmetries in the next step.
Our results suggest as well that anharmonicity cannot be neglected in any model or calculation trying to describe the free energies of the candidate low-symmetry phases, especially given that the Born-Oppenheimer energies of the competing phases are only few meV per formula unit apart \cite{ratcliff_2021}.

\subsection{The formation of the CDW phase}
Despite addressing anharmonicity as the primary factor in the CDW melting process, the mechanism responsible for the formation of the CDW in $\mathrm{CsV_{3}Sb_{5}}$ remains unclear.
In order to unveil this issue, we compute both the nesting function and the phonon linewidth given by the electron-phonon interaction along the $AL$ high-symmetry line, which exhibits the most unstable phonons.
These two quantities are very similar, with the difference that the latter includes the electron-phonon matrix elements but not the former as discussed in the supplementary material.
In Fig. \ref{fig:3}(a) we compare these two quantities for two modes: the mode responsible for driving the instability (labeled as $\alpha$) and the next most unstable mode (labeled as $\beta$).
The nesting function displays a prominent peak at the $A$ point, reflecting the highly two-dimensional Fermi surface of $\mathrm{CsV_{3}Sb_{5}}$ (Fig. \ref{fig:1}(b)), along with a smaller peak around $\frac{3}{4}$ $AL$, corresponding to the nesting vector between the Van Hove pockets at $k_{z}=0$ and the ones at $k_{z}=0.5$ (Fig. \ref{fig:1}(d)).
Conversely, the electron-phonon linewidth exhibits a significant increase from nearly zero at the $A$ point to a huge value of approximately $2\ \mathrm{meV}$ at the $L$ point for the $\alpha$ mode,  while remaining relatively constant for the $\beta$ mode.
Table S1 in the supplementary material provides an explicit comparison of the electron-phonon linewidths for all modes at the $M$ and $L$ points, highlighting the significant linewidths of the $L_2^{-}$ and $M_1^{+}$ modes compared to the others.
These findings further support the idea that the CDW is primarily mediated by the electron-phonon coupling rather than a nesting mechanism, underscoring once again the critical role of lattice effects in this system.

\begin{figure*}[]
		\centering
		\includegraphics[width=\linewidth]{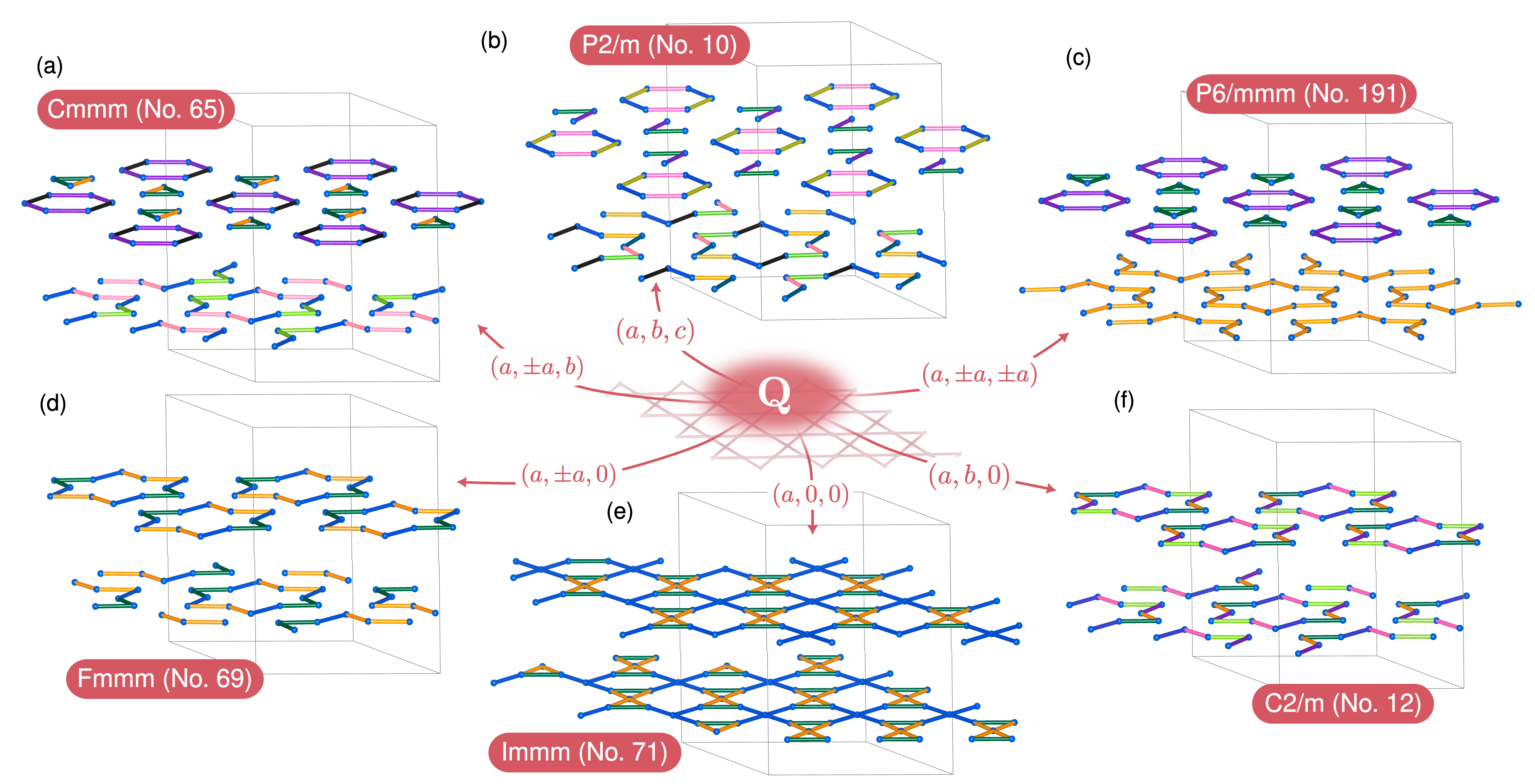}
		\caption{
		\textbf{All possible space groups compatible with the order parameter $\mathbf{Q}=(L_1,L_2,L_3)$.}
		Different colors distinguish between different bond lengths within the distorted kagome lattices of vanadium atoms.
		The most symmetric configuration, depicted in subfigure \textbf{c}, corresponds to what is usually referred in the literature as a stack of star-of-david and tri-hexagonal (inverse-star-of-david) configurations.
		The remaining distortions either represent distorted versions of \textbf{c} \textbf{(a,b)} or a stack of star-of-david configurations, but with the stars being deformed in some form \textbf{(d,e,f)}.
		}
		\label{fig:5}
\end{figure*}

\subsection{The anharmonic spectral function}
Our results indicate either a second-order or a weak first-order character for the phase transition, with the continuous nature imprinted in the softening of the $L_2^{- }$ phonon (Fig. \ref{fig:2}(b)).
However, multiple instances in the literature point to a first-order nature of the CDW \cite{ratcliff_2021,xiao_2023}, and neither inelastic x-ray scattering \cite{subires_2023} or Raman spectroscopy experiments  \cite{liu_2022a,he_2023} have observed such softening.
To understand this apparent contradiction, we compute the spectral function for both the $M$ and $L$ phonons fully accounting for anharmonic effects.
The spectral function, as shown in Fig. \ref{fig:4}(a,b), not only confirms the absence of phonon collapse at the $M$ point, but also underscores the profoundly anharmonic nature of the $L_2^{- }$ phonon that drives the instability.
As depicted in Fig. \ref{fig:4}(b), the unstable $L_2^{- }$ mode becomes broader and splits into a double peak as the temperature approaches $T_{CDW}$.
Thus, we expect this mode to be exceedingly challenging to observe experimentally.
This difficulty arises not only from its double-peak and broadened nature but also because the spectral function is typically fitted using a set of Lorentzians representing different modes experimentally.
To study the cumulative effects of both anharmonic and electron-phonon linewidths, we first need well-defined quasiparticles to which we can later add the electron-phonon linewidth.
To achieve this, we compute the spectral function using a no-mixing approximation by excluding the off-diagonal elements of the anharmonic phonon self-energy.
This allows for the independent tracking of different modes, as indicated by different colors in Fig. \ref{fig:4}(c,d).
Nonetheless, for illustrative purposes, the full anharmonic spectral functions are overlaid to demonstrate the impact of this approximation.
This approximation works effectively for the $M$ point, except for the softest mode at 50 K, and reveals the softening of a highly anharmonic mode (colored in blue) that corresponds to $M_{1}^{+}$.
In the case of \(L\), off-diagonal terms exert a more significant influence overall, as evidenced by the stronger differences between panels (b) and (d).
For instance, the softening in panel (d) appears at a lower temperature.
Still, we are able to capture the softening of a highly anharmonic mode (colored in blue), which develops a double dome and corresponds to $L_{2}^{-}$.
In Fig. \ref{fig:4}(e,f) we fit a Lorentzian function to each of the individual modes and add the corresponding electron-phonon linewidth in order to obtain a spectral function that captures both anharmonic and electron-phonon interactions.
The exact values for the centers and linewidths of the Lorentzian functions at temperatures of 200 K and 50 K are provided in Table S1 of the supplementary material.
Both at $M$ and $L$, the phonons highlighted in blue ($M_{1}^{+},L_{2}^{-}$) experience such a substantial broadening that they become imperceptible to the eye.
This suggests that the electron-phonon linewidth renders this mode experimentally inaccessible, implying that the softening goes unnoticed, reconciling our theoretical results with the experiments \cite{subires_2023,liu_2022a,he_2023}.
The remaining modes at $L$ are well-described by Lorentzian curves, exhibiting a slight broadening as the temperature increases, consistent with the expected impact of anharmonicity.

\subsection{Symmetry analysis of the potential resulting phases}
Based on our previous findings, we expect that the CDW transition manifests as either a Landau-type phase transition or a subtle first-order transition.
This, together with the observation that the $M_1^{+ }$ mode is stabilized by anharmonic effects, provides an opportunity to explore the potential phase transitions permitted by group theory \cite{stokes_1988}.
By independently varying the amplitudes of each component within the three-dimensional order parameter $\mathbf{Q}=\left(L_1,L_2,L_3 \right)$, we identify six distinct possible space groups.
Each of these distortions imprints a characteristic pattern between two adjacent kagome lattices as shown in Fig. \ref{fig:5}.
The resulting space groups are also predicted to exhibit characteristic Raman and infrared spectra, featuring varying numbers of peaks that transform under distinct symmetries (see Table S3).
The phonon responsible for the symmetry breaking is expected to transform under the trivial representation $A_{g}$ of the low-symmetry group.
This is because the symmetries that leave this phonon invariant are precisely the ones preserved in the low-symmetry phase.
Given that the identity or trivial representation is always contained in $[V]^2$ (the symmetrized square of the vector representation), the mode responsible for the symmetry breaking will always be Raman active in the low symmetry phase.
However, its observation will be hindered because of its large phonon broadening. 
These distinctions in Raman and infrared activities may offer a robust method for discerning the low-symmetry structure.
This stands in contrast to energetic arguments, which can be notably unreliable due to the minute energy differences between phases and the neglect of certain contributions, such as anharmonicity.

\section{Conclusions}
In summary, our calculations show that the CDW transition in $\mathrm{CsV_{3}Sb_{5}}$ is primarily driven by the large electron-phonon coupling within the system, while the melting of the CDW can be attributed to the robust anharmonic effects of the lattice.
The CDW is exclusively triggered by the unstable phonons at the $L$ point, with the $M$ phonons not assuming a pivotal role.
Despite the phase transition being of second order or weak first-order character, an examination of the spectral function at the $L$ point suggests that observing this softening experimentally is impossible \cite{subires_2023}.
All in all, in contrast to the pure kagome CDW, which is prototypically nesting driven and strictly two-dimensional, the $\mathrm{CsV_{3}Sb_{5}}$ CDW exhibits notable deviations being purely three-dimensional and driven by the electron-phonon coupling.
The remarkable agreement between our results and experimental data suggests that the CDW mechanism can be studied independently from the phenomena emerging in the CDW phase and that lattice quantum anharmonic effects may also wield a crucial influence on the novel physics in the CDW state.
We anticipate a similar impact of the electron-phonon coupling and anharmonicity on other kagome families \cite{korshunov_2023,li_2023a,teng_2023,dong_2023a}, as well as the presence of anomalous phonon spectral functions.
For instance, the discrepancy observed in 166 compounds between the phonon collapse and the CDW order \cite{korshunov_2023} might be attributed to the phonon responsible for the CDW order going unnoticed because of its large linewidth.
On the other hand, the absence of a CDW in titanium-based $\mathrm{CsTi_3Bi_5}$ kagome compounds \cite{li_2023a} likely results from a distinct balance between the stabilizing role of anharmonicity and electron-phonon destabilizing forces.

\section{Methods}
First-principles density functional theory (DFT) calculations were conducted using the Quantum Espresso package \cite{giannozzi_2009,giannozzi_2017}.
We used the generalized gradient approximation with the Perdew-Burke-Ernzerhof parameterization \cite{perdew_1996} together with projector-augmented wave pseudopotentials \cite{kresse_1999} generated by Dal Corso \cite{corso_2014} and considering $9 / 5 / 13$ valence electrons for cesium/antimony/vanadium.
Unless stated otherwise, we used energy cutoffs of 60/600 Ry for the wavefunctions/density with a Methfessel-Paxton smearing \cite{methfessel_1989} of 0.002 Ry for the calculations.
The structural relaxation and DFPT calculations were performed using a \(16 \times 16 \times 10\) grid, without accounting for spin-orbit coupling (SOC), and internal relaxations were done with the experimental lattice parameters given in \cite{ortiz_2019} and stopped when forces were below 0.001 Ry/au.
Subsequently, SOC was included to compute the electronic band structures and Fermi surfaces (see Fig. \ref{fig:1}).
The calculations using Fermi-Dirac smearing were done with an \(18 \times 18 \times 12\) \(k\)-grid and with smearings ranging from 0.0001 Ry up to 0.0065 Ry.
To generate the Fermi surface plots, we employed the Wannierization procedure implemented in Wannier90 \cite{mostofi_2014}, along with WannierTools \cite{wu_2018}.
First, we obtained a tight-binding model with a Wannierization considering d and p orbitals in vanadium and antimony sites.
Then, we calculated the Fermi surface as implemented in WannierTools for a $200 \times 2000 \times 100$ grid.
Harmonic phonons were computed using density functional perturbation theory (DFPT) \cite{baroni_2001} within a $6 \times 6 \times 4$ phonon grid.
The anharmonic temperature dependent phonon calculations were done under the Stochastic Self-Consistent Harmonic Approximation (SSCHA) \cite{errea_2014,bianco_2017,monacelli_2018} as implemented in the SSCHA code \cite{monacelli_2021}. 
In order to capture all the relevant high-symmetry points, the free energy Hessians (SSCHA anharmonic phonons) were done with the inclusion of fourth order force constants in a $2 \times 2 \times 2$ supercell, which naturally captures the $\Gamma$, $A$, $M$ and $L$ points.
The Brillouin zone integrals for the supercell calculations were performed with a $7\times 7\times 4$ $k$-grid
(equivalent to a $14 \times 14 \times 8$ grid in the primitive cell).
The dynamical extension of the theory \cite{bianco_2017,monacelli_2021a} was used to compute the spectral function within the so-called bubble approximation for the self-energy. 
The calculation considered phonon-phonon scattering on a $2\times 2\times 2$ and a $0.1\ \mathrm{cm^{-1}}$ Gaussian smearing was used to approximate the Dirac deltas.

\section{Data availability}
All relevant data are available from the authors upon reasonable request.

\section{Code availability}
All codes used in this study are open-source and available from their respective websites.

\begin{acknowledgments}
We acknowledge fruitful discussions with J. L. Ma\~nes. 
M.G.V., I.E. and M.G.A acknowledge the Spanish Ministerio de Ciencia e Innovaci\'on (grants PID2019-109905GB-C21, PID2022-142008NB-I00, and PID2022-142861NA-I00). I.E. acknowledges the Department of Education, Universities and Research of the Eusko Jaurlaritza and the University of the Basque Country UPV/EHU (Grant No. IT1527-22).  
M.G.A. thanks the Department of Education of the Basque Government for a predoctoral fellowship (Grant no. PRE\_2019\_1\_0304). 
M.G.V.  and C.F. thanks support to the Deutsche Forschungsgemeinschaft (DFG, German Research Foundation) GA 3314/1-1 – FOR 5249 (QUAST) and partial support from European Research Council (ERC) grant agreement no. 101020833. 
This work has also been  funded by the Ministry of Economic Affairs and Digital Transformation of the Spanish Government through the QUANTUM ENIA project call – Quantum Spain project, and by the European Union through the Recovery, Transformation and Resilience Plan – NextGenerationEU within the framework of the Digital Spain 2026 Agenda.
This project has received funding from the European Research Council (ERC) under the European Union’s Horizon 2020 research and innovation programme XBEND (Grant agreement No. 101080740).
\end{acknowledgments}

\section*{References}
\bibliography{bibtex.bib}

\section*{Author contributions}
I.E. and M.G.V. conceived and supervised the project, and also revised the manuscript.
M.G.-A. carried out the calculations and drafted the manuscript.
The results were discussed with C.G., P.J.W.M., and D.D., who also reviewed the manuscript and contributed suggestions.
C.F. and I.E. provided the computational resources.
All authors contributed to data analysis and reviewed the final version of the manuscript.

\section*{Competing interests}
The authors declare no competing interests.

\end{document}


\title{\emph{Supplementary Material for}\\
Phonon collapse and anharmonic melting of the 3D charge-density wave in kagome metals
}

\author{Martin Gutierrez-Amigo}
\affiliation{Donostia International Physics Center (DIPC), 20018 Donostia/San Sebasti\'an, Spain}
\affiliation{Department of Physics, University of the Basque Country (UPV/EHU), 48080 Bilbao, Spain}
\affiliation{Centro de Física de Materiales (CSIC-UPV/EHU), 20018 Donostia/San Sebasti\'an, Spain}

\author{{\DJ}or{\dj}e Dangi{\'c}}
\affiliation{Centro de Física de Materiales (CSIC-UPV/EHU), 20018 Donostia/San Sebasti\'an, Spain}
\affiliation{Fisika Aplikatua Saila, Gipuzkoako Ingeniaritza Eskola, University of the Basque Country (UPV/EHU), 20018 Donostia/San Sebasti\'an, Spain}

\author{Chunyu Guo}
\affiliation{Max Planck Institute for the Structure and Dynamics of Matter, 22761 Hamburg, Germany}

\author{Claudia Felser}
\affiliation{Max Planck Institute for Chemical Physics of Solids, 01187 Dresden, Germany}

\author{Philip J. W. Moll}
\affiliation{Max Planck Institute for the Structure and Dynamics of Matter, 22761 Hamburg, Germany}

\author{Maia G. Vergniory}
\affiliation{Max Planck Institute for Chemical Physics of Solids, 01187 Dresden, Germany}
\affiliation{Donostia International Physics Center (DIPC), 20018 Donostia/San Sebasti\'an, Spain}

\author{Ion Errea}
\affiliation{Centro de Física de Materiales (CSIC-UPV/EHU), 20018 Donostia/San Sebasti\'an, Spain}
\affiliation{Fisika Aplikatua Saila, Gipuzkoako Ingeniaritza Eskola, University of the Basque Country (UPV/EHU), 20018 Donostia/San Sebasti\'an, Spain}
\affiliation{Donostia International Physics Center (DIPC), 20018 Donostia/San Sebasti\'an, Spain}

\date{November 22, 2023}
\maketitle
\section*{Details of first-principle calculations}

\begin{figure*}[]
		\centering
		\includegraphics[width=0.75\textwidth]{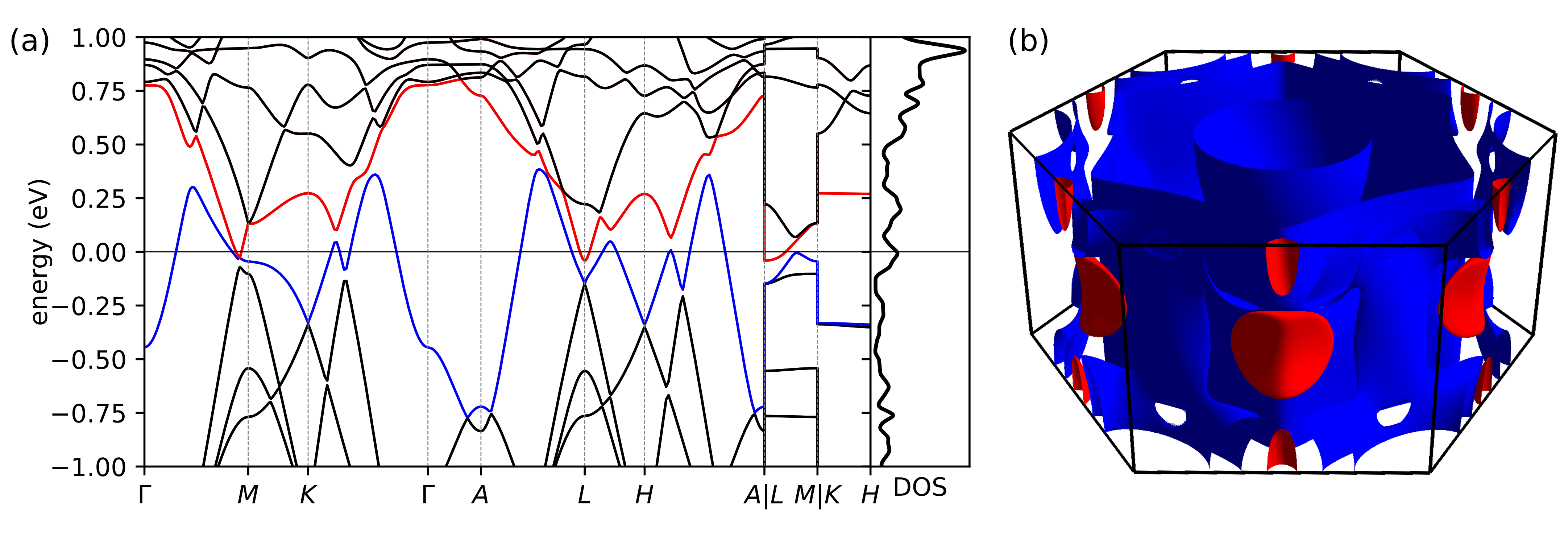}
		\caption{\textbf{Band structure and Fermi surface for $\mathrm{CsV_3Sb_5}$.}
		\textbf{a.} Band structure and density of states (DOS) with the two bands crossing the Fermi level highlighted in red and blue.
		\textbf{b.} Corresponding Fermi surface sections to the red/blue bands shown in panel \textbf{a}.
		}
		\label{fig:1a}
\end{figure*}

\begin{figure}[]
		\centering
		\includegraphics[width=\linewidth]{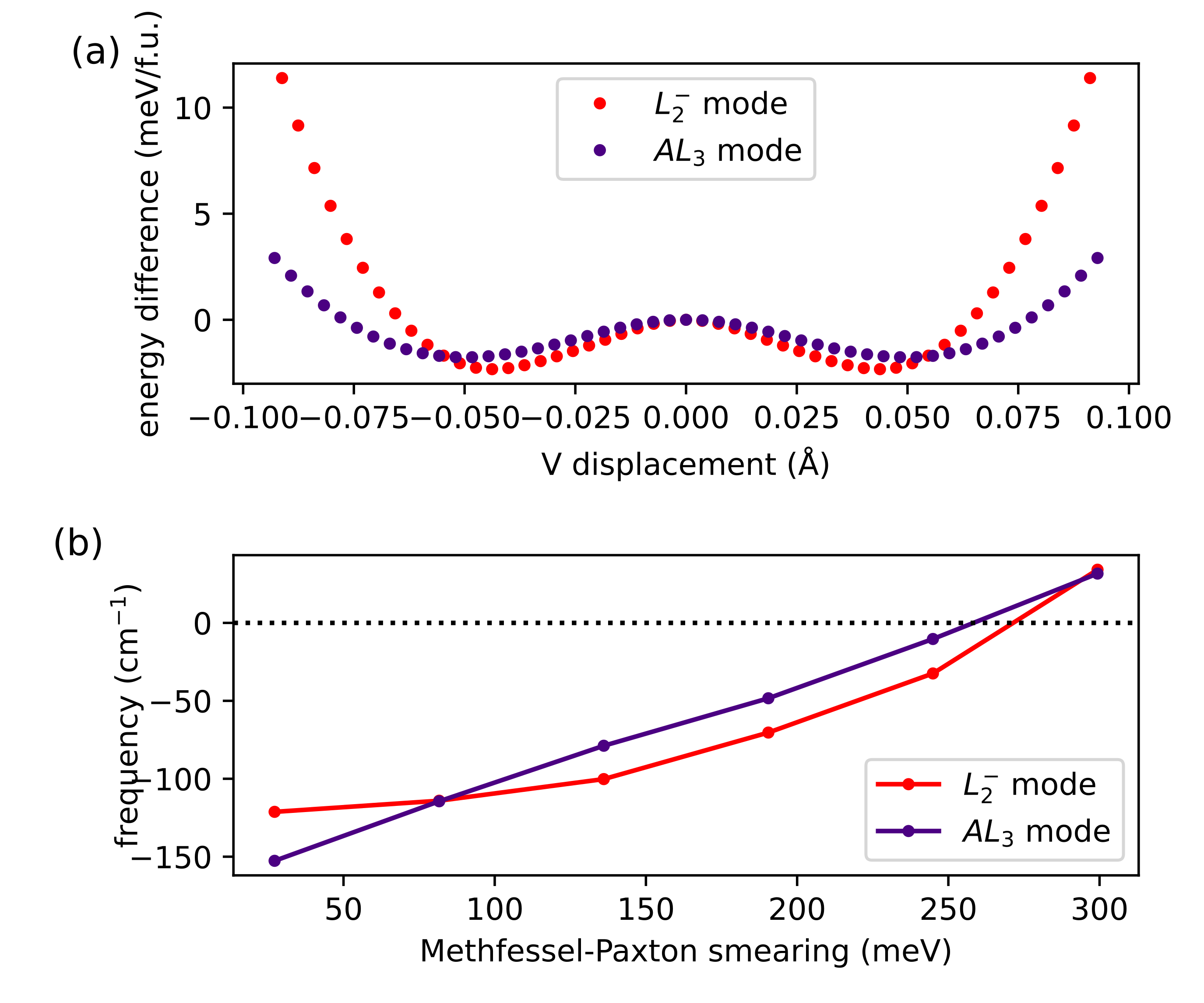}
		\caption{\textbf{Smearing dependence of $L$ and $AL$ modes.}
			\textbf{a.} Born-Oppenheimer energy surfaces as the crystal is distorted according to either the $L_{2^{- }}$ or the  $AL_{3}$ mode. As shown, the $L$ mode leads to a lower energy state.
			\textbf{b.} Frequency dependence of both $L_{2}^{-}$ and $AL_{3}$ modes as a function of a Methfessel-Paxton smearing.
		}
		\label{fig:1}
\end{figure}

\subsection*{Born-Oppenheimer energy surface calculation}
In order to compute the Born-Oppenheimer (BO) energy surface, we calculate the energy by displacing the ions according to the corresponding active phonons. 
In the harmonic approximation, the displacement $\mathbf{u}$ of an atom $s$ in the unit cell $\mathbf{R}$ can be expressed as:
\[
		{u}^{\alpha}_s(\mathbf{R})=\mathrm{Re}\{ \sum_{\mu \mathbf{k}} q_\mu (\mathbf{k}) \frac{\mathbf{\varepsilon}_{\mu s}^{\alpha}(\mathbf{k})}{\sqrt{M_s}}e^{i\mathbf{k}\cdot\mathbf{R}} \}
.\] 
Here, $\alpha$ is a Cartesian coordinate, $\mu$ labels the mode, $M_s$ represents the ionic mass of atom $s$, $\varepsilon_{\mu s}^{\alpha} (\mathbf{k})$ is the polarization vector, and $q_\mu(\mathbf{k})$ is the order parameter associated with the $\mu$ mode at wave number $\mathbf{k}$. 
Then, by plotting the energy against the order parameter $q$, we obtain the BO energy surface along that specific direction in the order parameter space.

\begin{figure*}[]
		\centering
		\includegraphics[width=\textwidth]{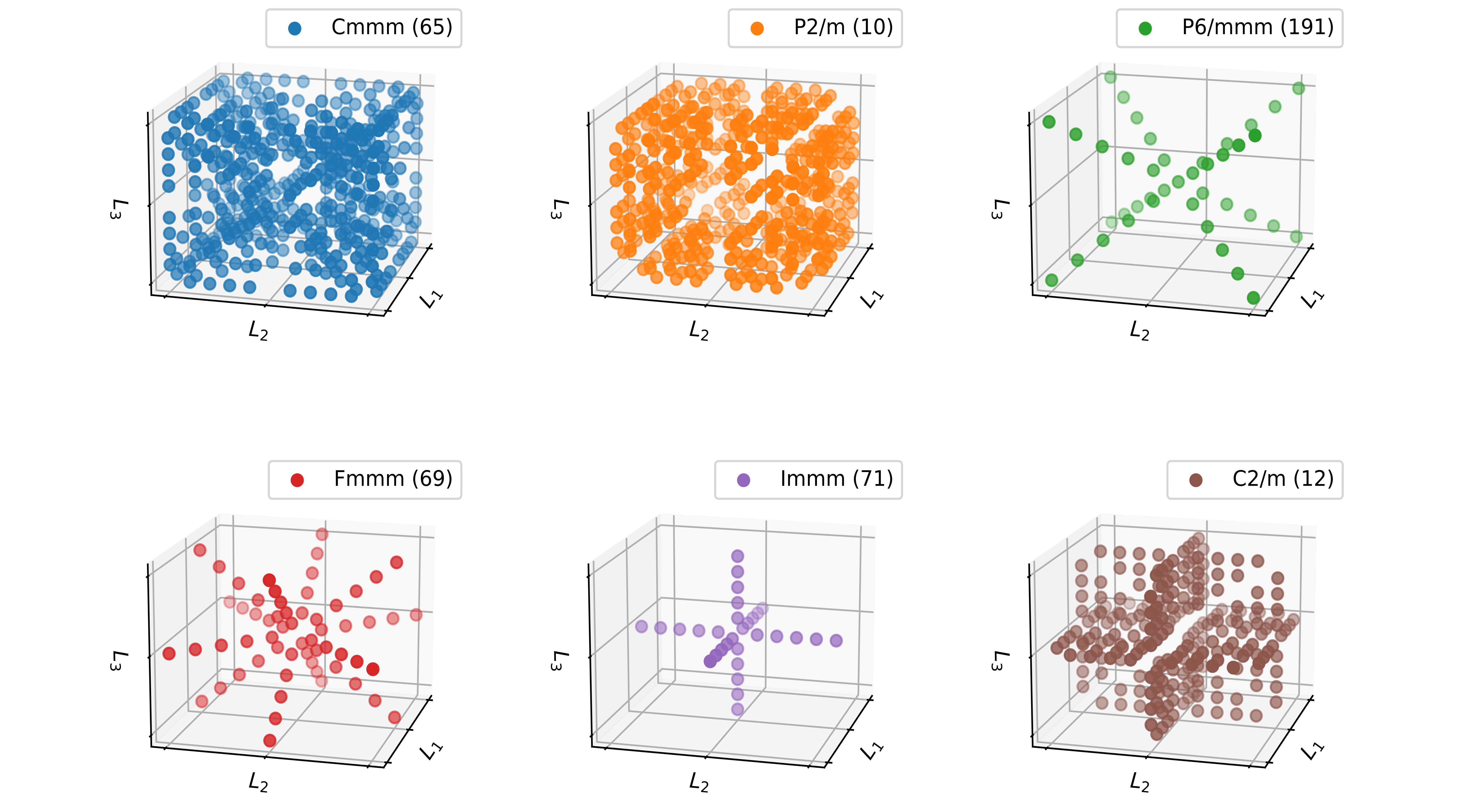}
		\caption{\textbf{Order parameter space for $\mathbf{Q}=(L_1,L_2,L_3)$.}
				Each direction in the three-dimensional order parameter space corresponds to a distinct instance of the distorted crystal, characterized by a specific space group and specific set of Wyckoff positions described in Table \ref{tab:1}.
				Here the different directions corresponding to each of the space groups are depicted.
		}
		\label{fig:2}
\end{figure*}

\subsection*{Electron-phonon linewidth and nesting function}
\begin{table*}[]
\centering
\begin{ruledtabular}
\begin{tabular}{ccccccccc|ccccccccc}
		\multicolumn{9}{c|}{$M$ mode} & \multicolumn{9}{c}{$L$ mode} \\ 
		Irrep & $\omega^{\mathrm{H}}$ & $\omega^{\mathrm{A}}_{200}$ & $\omega^{\mathrm{A}}_{50}$ & $\omega^{\mathrm{L}}_{200}$ & $\omega^{\mathrm{L}}_{50}$ & $\gamma^{\mathrm{A}}_{200}$ & $\gamma^{\mathrm{A}}_{50}$ & $\gamma^{\mathrm{e-ph}}$ 
		& Irrep & $\omega^{\mathrm{H}}$ & $\omega^{\mathrm{A}}_{200}$ & $\omega^{\mathrm{A}}_{50}$ & $\omega^{\mathrm{L}}_{200}$ & $\omega^{\mathrm{L}}_{50}$ & $\gamma^{\mathrm{A}}_{200}$ & $\gamma^{\mathrm{A}}_{50}$ & $\gamma^{\mathrm{e-ph}}$ \\ \hline
	$A_{g} $ &-89 &60 &47 &66 &36 &2.83 &0.59 &17.49	&$B_{1u}$ &-121 &57 &-32 &68 &24 &3.17 &2.21 &14.38	\\
	$B_{1u}$ &38 &38 &23 &38 &39 &1.15 &0.24 &0.06		&$B_{1u}$ &43 &39 &47 &40 &36 &1.04 &0.51 &0.09	\\
	$B_{2g}$ &42 &47 &47 &47 &47 &0.54 &0.18 &0.45		&$B_{2u}$ &43 &44 &43 &44 &44 &0.8 &0.15 &0.01	\\
	$B_{2u}$ &43 &43 &39 &43 &44 &0.82 &0.19 &0.01		&$B_{3u}$ &45 &46 &44 &46 &47 &0.66 &0.16 &0.22	\\
	$B_{3u}$ &46 &46 &44 &47 &47 &0.52 &0.17 &0.01		&$B_{3u}$ &49 &50 &51 &50 &51 &1.37 &0.29 &0.19	\\
	$B_{1g}$ &56 &55 &56 &56 &56 &0.61 &0.16 &0.03		&$A_{u }$ &56 &54 &56 &54 &56 &0.87 &0.02 &0.04	\\
	$B_{1u}$ &63 &56 &62 &61 &63 &1.63 &0.42 &0.07		&$A_{g }$ &60 &57 &60 &57 &60 &1.01 &0.32 &0.15	\\
	$A_{u }$ &79 &77 &78 &77 &79 &0.64 &0.19 &0.06		&$B_{1g}$ &79 &76 &78 &77 &78 &0.8 &0.31 &0.08	\\
	$B_{1u}$ &92 &87 &91 &90 &91 &1.24 &0.24 &0.29		&$A_{g }$ &90 &84 &90 &89 &91 &1.87 &0.84 &0.26	\\
	$B_{3u}$ &93 &89 &91 &88 &91 &1.61 &0.33 &0.13		&$B_{2g}$ &92 &87 &90 &85 &89 &2.04 &0.55 &0.07	\\
	$B_{3g}$ &104 &101 &104 &104 &104 &2.3 &0.6 &0.04	&$B_{3g}$ &103 &101 &103 &104 &104 &2.63 &1.3 &0.35	\\
	$B_{2u}$ &105 &101 &104 &101 &105 &4.11 &1.06 &0.25	&$B_{2u}$ &105 &103 &105 &104 &105 &0.82 &0.33 &0.05		\\
	$B_{2g}$ &113 &110 &112 &111 &112 &1.09 &0.73 &0.13	&$B_{3u}$ &115 &111 &115 &113 &115 &2.07 &0.16 &0.1 		\\
	$A_{g }$ &117 &112 &113 &115 &118 &1.67 &2.85 &1.46	&$B_{1u}$ &120 &113 &116 &117 &120 &3.03 &1.24 &0.42		\\
	$B_{3u}$ &122 &119 &122 &120 &121 &1.6 &3.97 &0.09	&$B_{2g}$ &122 &118 &121 &117 &120 &1.06 &1.72 &0.09		\\
	$B_{1u}$ &126 &122 &126 &123 &125 &3.49 &0.65 &0.28	&$A_{g }$ &126 &120 &125 &121 &125 &1.81 &0.44 &0.39		\\
	$B_{2u}$ &132 &127 &131 &129 &132 &3.32 &0.43 &0.13	&$B_{3g}$ &132 &126 &131 &127 &132 &2.22 &0.81 &0.13		\\
	$A_{g }$ &158 &165 &169 &174 &167 &11.68 &3.61 &4.43	&$B_{1u}$ &156 &165 &161 &175 &167 &10.3 &7.68 &2.45		\\
	$B_{3u}$ &161 &159 &160 &158 &160 &2.82 &0.83 &0.13	&$B_{2g}$ &161 &157 &159 &158 &160 &5.82 &0.71 &0.14		\\
	$B_{1g}$ &165 &159 &160 &168 &168 &3.23 &1.17 &0.91	&$A_{u }$ &167 &159 &169 &166 &169 &5.6 &1.47 &1.01 		\\
	$B_{3u}$ &207 &201 &208 &214 &210 &2.76 &1.87 &1.08	&$B_{2g}$ &208 &206 &207 &212 &208 &2.35 &1.04 &1.02		\\
	$A_{g }$ &211 &212 &216 &219 &216 &6.37 &0.81 &3.4	&$B_{1u}$ &215 &216 &217 &219 &216 &3.37 &1.13 &1.84		\\
	$B_{1g}$ &216 &211 &218 &216 &217 &6.48 &0.83 &0.6	&$A_{u }$ &216 &214 &218 &215 &218 &1.61 &0.45 &0.7 		\\
	$B_{2u}$ &244 &244 &245 &249 &245 &4.47 &1.22 &2.59	&$B_{3g}$ &242 &246 &244 &250 &245 &2.47 &1.0 &2.53 		\\
	$B_{2g}$ &258 &251 &257 &257 &258 &6.66 &1.96 &1.78	&$B_{3u}$ &256 &249 &257 &256 &258 &4.26 &2.12 &1.58		\\
	$B_{1u}$ &261 &257 &260 &262 &261 &4.06 &1.59 &1.48	&$A_{g }$ &260 &254 &259 &259 &260 &3.24 &2.07 &2.04		\\
	$B_{3g}$ &269 &265 &273 &268 &272 &5.26 &1.58 &4.63 	&$B_{2u}$ &277 &269 &273 &271 &273 &3.16 &1.47 &5.93	
\end{tabular}
\end{ruledtabular}
\caption{
\textbf{Frequencies and phonon-phonon and electron-phonon linewidths for all $M$ and $L$ modes.}
First, we provide the irreducible representation for each of the modes.
Then we provide the frequencies of the mode under the harmonic approximation ($\omega^{\mathrm{H}}$), as the Hessian of the free energy ($\omega^{\mathrm{A}}$) at 200 K and 50 K (see Fig. 2 (a,b)), and as the center of the Lorentzian fits of the spectral function computed under the no-mixing approximation in Fig. 4 (c,d), also at 200 K and 50 K.
Lastly, we provide the electron-phonon ($\gamma^{\mathrm{e-ph}}$) and phonon-phonon ($\gamma^{\mathrm{A}}$) HWHM linewidths, with the latter also given for 200 K and 50 K.
The units for all results are in $\mathrm{cm^{-1}}$.
}
\label{tab:1a}
\end{table*}

The electron-phonon matrix elements $g_{n\mathbf{k},m\mathbf{k+q}}^{\mu}$ for a phonon mode $\mu$ with momentum $\mathbf{q}$ and two electronic states in bands $n$ and $m$ with electronic momenta $\mathbf{k}$ and $\mathbf{k+q}$ are calculated withing DFPT as:
\[
g_{n\mathbf{k},m\mathbf{k}+\mathbf{q}}^{\mu}=\sum_{s\alpha}
\frac{1}{\sqrt{ 2M_{s}\omega_{\mu}(\mathbf{q}) }}\varepsilon _{\mu s}^{\alpha}
\braket{ n\mathbf{k} | \left[ \frac{ \partial V_{KS} }{ \partial u_{s}^{\alpha}(\mathbf{q}) } \right]_{0}  |m\mathbf{k}+\mathbf{q} } 
.\] 
where $M_{s}$ is the atomic mass of atom $s$, $\omega_{\mu}(\mathbf{q})$ is the frequency of the mode, $\varepsilon_{\mu s}^{\alpha}$ is the polarization vector, with $\alpha$ being a Cartesian direction and $\braket{ n\mathbf{k} | \left[ \frac{ \partial V_{KS} }{ \partial u_{s}^{\alpha}(\mathbf{q}) } \right]_{0}  |m\mathbf{k}+\mathbf{q} } $ are the matrix elements of the derivative of the Kohn-Sham potential with respect to the atomic displacements of the phonon mode.
Then, the electron-phonon contribution to the phonon half-width-half-maximum (HWHM) linewidth for mode $\mu$ with momentum  $\mathbf{q}$ can be calculated as:
\[
HWHM_{elph,\mu}(\mathbf{q})=\frac{2 \pi \omega_{\mu}(\mathbf{q})}{N_{k}}\sum_{\mathbf{k}n m}
\left| g_{n\mathbf{k},m\mathbf{k+q}}^{\mu} \right|^{2} \delta(\epsilon _{n\mathbf{k}})\delta(\epsilon _{m\mathbf{k+q}})
,\] 
with $N_{k}$ being the number of $\mathbf{k}$ points in the sum and $\epsilon_{n\mathbf{k}}$ the energy of the state $\ket{n\mathbf{k}}$ measured from the Fermi level. Notice that $HWHM_{elph,\mu}$ is independent from the frequency $\omega_{\mu}(\mathbf{q})$, since this term cancels with the one from $g_{n\mathbf{k},m\mathbf{k+q}}^{\mu}$, which allows us to define it even for negative frequencies.
The $HWHM_{elph,\mu}$ was computed using a $24 \times\ 24 \times\ 15$ grid and a Gaussian smearing of 0.005 Ry for the Dirac deltas.
The results for all modes at the $M$ and $L$ points are provided in Table \ref{tab:1a}, with the $M_{1}^{+}$, $L_{2}^{-}$ and $\beta$ modes discussed in the main text corresponding in the first two rows.
The nesting function $\zeta(\mathbf{q})$ was computed using the same grid and Gaussian smearing and defined as:
\[
\zeta\left( \mathbf{q}\right)=\frac{1}{N_{k}} \sum_{\mathbf{k}nm}\delta(\epsilon _{n\mathbf{k}})\delta(\epsilon _{m\mathbf{k+q}})
.\]

\begin{table*}[]
\begin{center}
\begin{ruledtabular}
\begin{tabular}{ccccc}
Space Group & Order parameter & Cesium & Antimony & Vanadium \\ \hline
$P6/mmm$ (No. 191)   &  $(0,0,0)$   & $1a$ & $1b+4h$ & $3g$ \\
$P6/mmm$ (No. 191)   & $(a,\pm a,\pm a)$   & $2e+6i  $ & $3f+ 3g+ 1a+ 1b+ 2\times (12o+ 4h)$ & $6l+ 6j+ 6m+ 6k$ \\
$Immm$  (No. 71)     & $(a,0,0)$     & $4i$ & $2a+ 2c+ 2\times 8l$ & $2b+ 2d+ 8n$ \\
$Fmmm$ (No. 69)     & $(a,\pm a,0)$    &$8f+ 8h$ &$4a + 8d+ 32p + 4b+ 2\times 16m $ & $8i+ 8g+ 2\times 16n$\\
$Cmmm$ (No. 65)& $(a,\pm a,b)$   &$4k+ 4l+ 8m$  &\makecell{$2a+ 2b+ 2d+ 2c+ 4e+ 4f+$\\$ + 4\times 8n + 2\times 16r$}& $4g+ 4h+ 4i+ 4j+ 2\times (8p + 8q)$ \\
$C2/m$ (No. 12)& $(a,b,0)$ &$4g+ 4h$ &$2a+ 2b+ 2c+ 2d+ 4\times 8j$ & $6\times 4i$ \\
$P2/m$ (No. 10)& $(a,b,c)$ & $2i+ 2k+ 2j+ 2l$ & \makecell{$1a+ 1b+ 1c+ 1d+ 1e+ 1f+$\\$+1g+ 1h+ 8\times 4o$} & $6\times (2m+ 2n)$
\end{tabular}
\end{ruledtabular}
\end{center}
\caption{
\textbf{Wyckoff positions for all possible subgroups resulting from the condensation of the $L_2^{-}$ modes.}
The first column indicates the resulting space group corresponding to the order parameter $\mathbf{Q}=(L_1,L_2,L_3)$ in the second column.
The third fourth and fifth columns specify the Wyckoff positions for cesium, antimony and vanadium atoms in each of the space groups.
}
\label{tab:1}
\end{table*}

\begin{table*}[]
\centering
\begin{ruledtabular}
\begin{tabular}{ccccc}
Space Group & Order parameter & Num. modes & Raman active & Infrared active \\ \hline
$P6/mmm$ (No. 191)   &  $(0,0,0)$      & 27	& $A_{1g}+E_{2g}+E_{1g}$ & $4A_{2u}+5E_{1u}$   \\
$P6/mmm$ (No. 191)   & $(a,\pm a,\pm a)$   & 216 &$12 A_{1g}+17E_{2g}+15E_{1g}$  &$16 A_{2u}+25E_{1u}$ \\
$Immm$  (No. 71)     & $(a,0,0)$       & 54	& $7A_{g}+4B_{1g}+4B_{2g}+6B_{3g}$ & $10B_{1u}+11B_{2u}+9B_{3u}$ \\ 
$Fmmm$ (No. 69)     & $(a,\pm a,0)$     & 108 & $14 A_{g}+10B_{1g}+12B_{2g}+12B_{3g}$ & $\ 19B_{1u}+16B_{2u}+17B_{3u}$ \\
$Cmmm$ (No. 65)& $(a,\pm a,b)$     & 216 & $29 A_{g}+23B_{1g}+20B_{2g}+24B_{3g}$ & $31B_{1u}+38B_{2u}+34B_{3u}$\\
$C2/m$ (No. 12)& $(a,b,0)$       & 108	& $26 A_{g}+22B_{g}$ & $24A_{u}+36B_{u}$ \\
$P2/m$ (No. 10)& $(a,b,c)$       & 216	& $52A_{g}+44B_{g}$ & $48A_{u}+72B_{u}$ 
\end{tabular}
\end{ruledtabular}
\caption{
\textbf{Raman and infrared active modes for all possible subgroups resulting from the condensation of the $L_2^{-}$ modes.}
The first column indicates the resulting space group corresponding to the order parameter $\mathbf{Q}=(L_1,L_2,L_3)$ in the second column.
The third column denotes the total number of modes in $\Gamma$, while the fourth and fifth columns provide the counts of Raman and infrared active modes along with their respective irreducible representations.
}
\label{tab:2}
\end{table*}

\subsection*{Anharmonic phonon spectra and spectral function}
In order to assign a particular Lorentzian to each mode we used the no-mixing approximation, consisting on discarding the off-diagonal elements in the computation of the self-energy.
Then, the corresponding spectral functions of each of the modes were fitted by a Lorentzian function using the least squares method.
Using the polarization vectors we were able to match the different modes and corresponding linewidths within different temperatures and with the electron-phonon linewidths calculations.
The explicit results for all modes at the $M$ and $L$ points at temperatures of 200 K and 50 K are provided in Table \ref{tab:1a}.

\section*{Comparison between L and AL modes}
According to our calculation of the harmonic phonon spectrum the most prominent instability occurs at a specific point along the $AL$ line, which we will refer to as the $AL$ mode.
However, the instabilities in the phonon spectra of $\mathrm{CsV_{3}Sb_{5}}$ are highly sensitive to the electronic temperature, as previously noted in \cite{christensen_2021}.
It is important to note that the electronic temperature commonly referred to in ab-initio calculations is merely a means to smear out the density of states for faster convergence \cite{methfessel_1989}, maintaining a relation with the electronic temperature, but without a one-to-one mapping.
A more meticulous analysis of the comparison between the $L$ and $AL$ instabilities is presented in Fig. \ref{fig:1}(a), where we studied the Born-Oppenheimer energy landscape by distorting the structure according to both $L$ and $AL$ modulations independently.
The calculations were performed with a smearing of 0.002 Ry (27 meV) and grids of \(16 \times 8 \times 5\) for the \(L\) mode and \(16 \times 6 \times 5\) for the \(AL\) mode.
As shown in Fig. \ref{fig:1}(a), both profiles exhibit a highly anharmonic “Mexican-hat” shape, with the \(L\) phonon leading to a lower-energy state.
It is worth mentioning that, in this case, the DFPT method and the frozen phonon approach for computing phonon frequencies do not yield the same result, highlighting the challenge of accurately addressing this unstable mode in the limit of very small smearings.
Additionally, Fig. \ref{fig:1}(b) shows the strong dependence on smearing for both instabilities, indicating that depending on the smearing value, either one could appear as the leading parameter for the charge-density wave.
However, it is observed that the $L$ mode stabilizes at higher temperatures, aligning with the findings in existing literature \cite{tan_2021,subedi_2022,wang_2022}.

\section*{Symmetry analysis and Born-Oppenheimer energy surface calculation}
In order to deduce the possible low symmetry structures we explore the possible space groups that span from the three-dimensional order parameter $\mathbf{Q}=(L_1,L_2,L_3)$.
We found that six possible space-groups arise for different kinds of linear combinations of the order parameter as depicted in Fig. \ref{fig:2} and represented in Fig. 5 of the main text.
In order to obtain the Raman and infrared activities of each of this phases we distort the structure according to each of the order parameters and solve the Wyckoff positions of the resulting structures in Table \ref{tab:1}.
Then, one can build the corresponding mechanical representation and decompose it into irreducible representations in the $\Gamma$ point. 
The vibrational modes will be infrared or Raman active depending on whether the corresponding irreducible representations are contained in the vector representation $V$ or its symmetrized square $[V]^2$ respectively (see Table \ref{tab:2}).
Given that the majority of modes do not undergo any softening during the CDW and that the active modes come from either $\Gamma$, $A$, $M$ or $L$ points, a more detailed classification of the resulting frequencies for each Raman and infrared mode could be performed.

\bibliographystyle{apsrev4-1} %
\bibliography{bibtex.bib}